\documentclass[reqno,prl,showpacs,twocolumn,nofootinbib]{revtex4-1}

\usepackage[centertags]{amsmath}
\usepackage{amsfonts}
\usepackage{amssymb}
\usepackage{amsthm}
\usepackage{newlfont}
\usepackage{stmaryrd}
\usepackage{mathrsfs}
\usepackage{euscript}
\usepackage{graphicx}
\usepackage{enumerate}
\usepackage{tikz}
\usepackage{pgf,ulem}
\usetikzlibrary{positioning,fit,calc}
\usetikzlibrary{arrows,automata}
\usepackage{wrapfig}
\usepackage{subfigure}
\usepackage{amscd}
\usepackage{hyperref}

\theoremstyle{plain}

\theoremstyle{definition}

\theoremstyle{remark}

 \let\be=\beta  
\let\ve=\varepsilon   
 \let\la=\lambda

\DeclareMathAlphabet{\mathpzc}{OT1}{pzc}{m}{it}

\newcommand{\id}{\textrm{d}}

\def\dbar{\,{\mathchar'26\mkern-12mu \text{d}}}

\def\bea{\begin{eqnarray}}
\def\eea{\end{eqnarray}}
\def\ba{\begin{array}}
\def\ea{\end{array}}

\def\la{\langle}
\def\ra{\rangle}

\begin{document}

\title{How statistical forces depend on thermodynamics and kinetics of driven media}

\author{Urna Basu, Christian Maes}
\affiliation{Instituut voor Theoretische Fysica, KU Leuven, Belgium}
\author{Karel Neto\v{c}n\'{y}}
\email{netocny@fzu.cz}
\affiliation{Institute of Physics, Academy of Sciences of the Czech Republic, Prague, Czech Republic}

\begin{abstract}
  We study the statistical force of a nonequilibrium environment on a quasi-static probe.  In the linear regime the isothermal work on the probe equals the excess work for the medium to relax to its new steady condition with displaced probe.  Also the relative importance of reaction paths can be measured via statistical forces, and from second order onwards the force on the probe reveals information about nonequilibrium changes in the reactivity of the medium.  We also show that statistical forces for nonequilibrium media are generally nonadditive, in contrast with the equilibrium situation. Both the presence of non-thermodynamic corrections to the forces and their nonadditivity put serious constraints on any formulation of nonequilibrium steady state thermodynamics.
\end{abstract}

\maketitle

Statistical forces appear in macroscopic physics as gradients of thermodynamic potentials \cite{KardarRMP}.  The equilibrium pressure of a gas in a container is for example given in terms of the change in free energy under a small extension of the volume, an essential tool for obtaining the equation of state.  In irreversible thermodynamics such statistical forces translate the typical tendency of macroscopic systems to reach equilibrium and generate fluxes of matter, momentum or energy \cite{Groot}.  Another simple and speaking example is that of the entropic force in a spring  as governs basic polymer physics \cite{entrspring}.

It is of much current interest to extend the study of statistical forces to nonequilibrium regimes \cite{Sengers,Kafri15,kafri14a}.  There are multiple motivations beyond the general ambition of constructing a nonequilibrium statistical mechanics. Understanding the nature of pressure in  active media or the change in osmotic pressure in presence of active solutes  is fundamental for the viability of steady state thermodynamics \cite{kafri14b,sasatasaki,osmosis}. There are also a growing number of practical aspects of using nonequilibrium environments to modify the nature of statistical forces \cite{aron}.  For example, it is important to explore the possible phenomenology of Casimir forces \cite{kardar} or of forces derived from vector and higher order potentials that induce oscillatory behavior on the probe's motion.   Finally, and as we show, statistical forces give better operational meaning to the notion of dynamical activity as recently explored for nonequilibrium purposes \cite{fren,vW,chan}.

This Letter is a systematic theoretical account on the meaning of statistical forces within the standard set-up of classical mesoscopic systems.  Our main findings are first that the linear order correction to the equilibrium statistical force is still of thermodynamic nature (but not in the form of a gradient).  In that regime statistical forces measure excess work which is a key-quantity in present accounts of steady--state thermodynamics \cite{sasatasaki,Oono}. Secondly, we find that statistical forces can measure dynamical activity.  That includes both measuring the relative activity through different reaction paths, and specifying how the external driving affects reactivities.  Thirdly we discuss why statistical forces outside equilibrium become nonadditive.  That connects with typical long range effects in nonequilibrium.

All of these aspects yield basic understanding of statistical forces outside equilibrium and can already be illustrated with toy models of mesoscopic systems that catch the essential physics.

\begin{figure}[ht]
  \centering
  \includegraphics[width=4 cm]{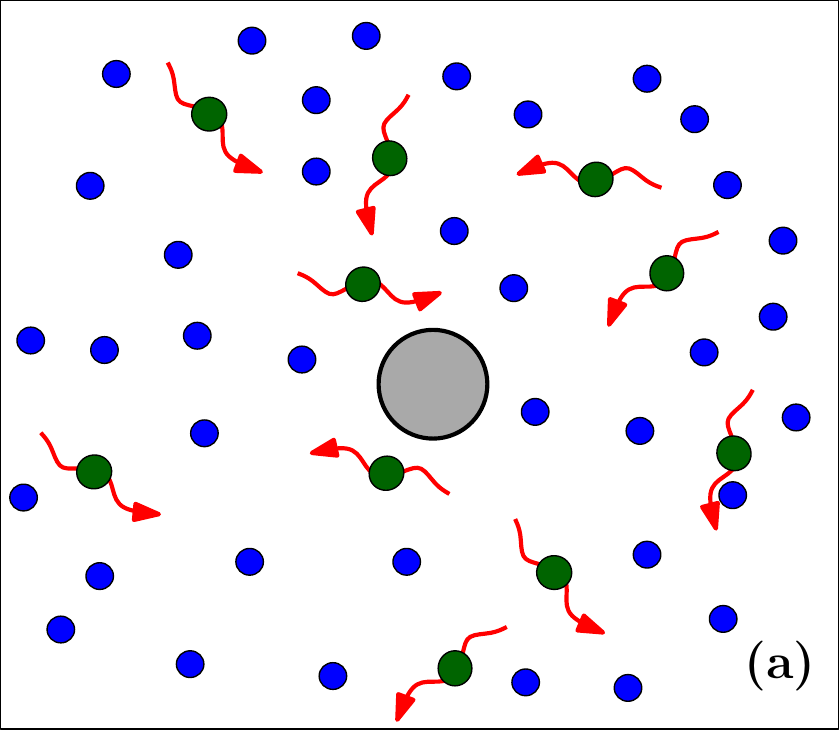}\hspace*{0.3 cm} \includegraphics[width=3.9 cm]{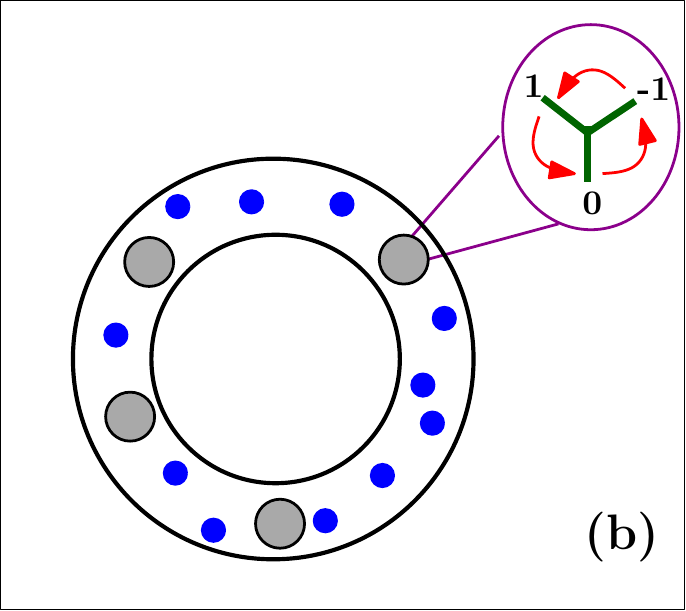}
  \caption{(Color online) (a) A slow probe (light gray disc) is immersed in a
  nonequilibrium medium (dark green circles), in contact with an equilibrium reservoir (small blue circles). (b) A dilute colloid solution diffusing on a ring, each colloid particle being coupled to an internal degree of freedom which is driven out of equilibrium. }
  \label{fig:setup}
\end{figure}

\noindent {\bf Set-up:}
Imagine  a nonequilibrium medium with a family $\eta$ of driven degrees of freedom.  It for example consists of particles undergoing nonconservative forces dissipating into a thermal bath at inverse temperature $\beta$.  A slow probe with position $x$ (a possibly high-dimensional or even collective coordinate) is immersed in the medium, with interaction potential $U(x,\eta),$ which is assumed also to include the self-interaction of the medium. Fig.~\ref{fig:setup}(a) shows a representative cartoon of the set-up. We  study the statistical force
\[
f(x) = -\int \rho_x(\id \eta) \, \nabla_x U(x,\eta) = -\langle \nabla_x U(x,\eta) \rangle^x
\]
where the average is over the steady nonequilibrium statistical distribution $\rho_x$ of the $\eta-$medium at fixed $x$. We assume there that the variables $\eta$ are relaxing very fast on the typical time-scale of motion of the probe position.  In other words, the probe moves in a medium which is in instantaneous stationarity, be it nonequilibrium. Note that the above formalism employs Newton's action--reaction law between probe and medium.

When the medium dynamics satisfies detailed balance, the statistical force $f_\text{eq}(x)= -\nabla_x {\cal F}(x)$ is given in terms of the free energy ${\cal F}(x) = -\frac 1{\beta} \log Z_x$ with the equilibrium partition function $Z_x$ corresponding to the $\eta$-medium when in equilibrium with fixed probe position $x$.  The ensuing motion of an overdamped probe would be gradient flow along $\dot{x} = -\nabla_x {\cal F}(x)$
(on the appropriate slow time scale)
to reach equilibrium at the position $x$ that minimizes the free energy.

Before discussing the thermodynamic and kinetic aspects  of statistical force from a driven medium in more detail  we illustrate the set--up and issues with the help of a simple example of a diffusion coupled to nonequilibrium degrees of freedom.\\

{\it Example---diffusion coupled to rotator}---The probe is a colloid trapped on the unit circle, $x\in S^1$, in contact with a thermal bath at inverse temperature $\beta=1$, modeled via the overdamped Langevin dynamics
\bea
\gamma\dot{x} = -\frac{\partial}{\partial x} U(x,\eta) + \sqrt{2\gamma \over \beta}~ \xi_t
\eea
where $\gamma$ is the damping and  $\xi_t$ is standard white noise. It is coupled to an `internal' degree of freedom $\eta=-1,0,1$ (Fig.~\ref{fig:setup}(b)) through the potential
\bea
U(x,\eta) = \eta \sin x + 2 \eta^2 \cos x \label{eq:Ux}
\eea
The choice  of potential is rather arbitrary but avoids special symmetries.
The fast degree of freedom $\eta$ is driven and rotates with transition rates
\bea\label{aa}
k^x(\eta,\eta')= e^{-\frac{\beta}2 [U(x,\eta')-U(x,\eta)]} \,\phi(\eta,\eta')\, e^{\frac{1}{2} s(\eta,\eta')}
\eea
The drive affects both the antisymmetric $s(\eta,\eta')=-s(\eta',\eta)$
and symmetric $\phi(\eta,\eta') = \phi(\eta',\eta)$ parts of the rates. We consider a uniform drive with field $\varepsilon$,
$s(-1,1)= s(1,0)= s(0,-1) = \beta \varepsilon$ and for simplicity we assume that only one of the reactivities gets changed by the drive
$\phi(-1,1) = \phi(1,-1)= \phi_0(1 + a |\varepsilon|)$ for some $a>0$, while
$\phi(0,\pm 1)=\phi(\pm 1,0)=1$.  The equilibrium (detailed balance) reference has $\ve=0$.  There, the $\phi_0$ picks up the relative importance of the dynamical activity over the transitions $1\leftrightarrow -1$.

\begin{figure}[thb]
  \centering
  \includegraphics[width=8.8cm]{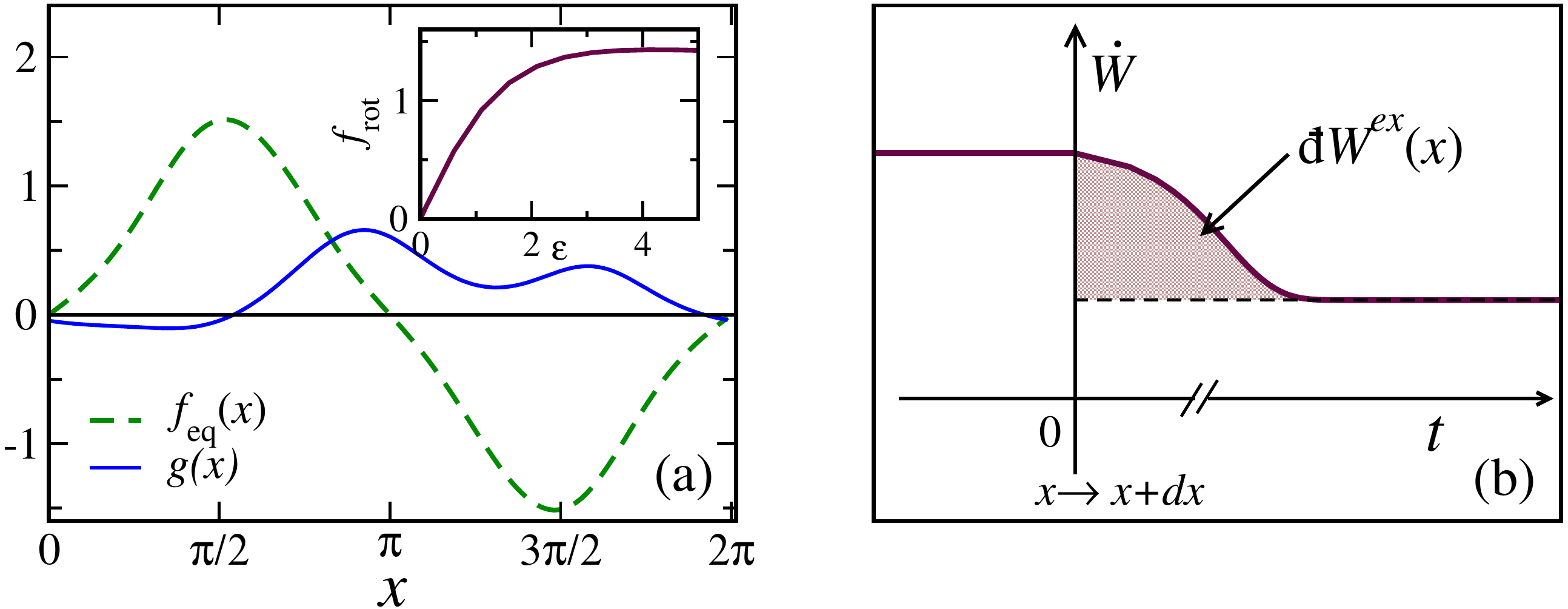}
  \caption{(Color online) (a) The equilibrium statistical force $f_\text{eq}(x)$
  and the nonequilibrium correction to it $g(x)$ for field $\varepsilon=2.0.$ Here $\phi_0=1,a=2.$ The inset shows the rotational part of the force $f_\text{rot}$ as a function of the field $\varepsilon.$
  (b) The excess work $ \dbar W^\text{ex}(x)$ is the extra dissipated work for relaxing to the new steady nonequilibrium (at different power $\dot W$) as the probe position $x\rightarrow x+\id x$ has changed.} 
  \label{fig:force}
\end{figure}

The statistical force on the particle is
\begin{equation}
f(x) = -\la \eta \ra^x \cos x + 2 \la \eta^2 \ra^x \sin x
\end{equation}
The equilibrium force $f_\text{eq}(x)$
and nonequilibrium correction
$g(x) = f(x) - f_\text{eq}(x)$
are plotted in Fig.~\ref{fig:force}(a).
While the force is derived from a potential in equilibrium that is no longer true when the $\eta$ are driven; the rotational part of the force
$f_\text{rot} = \oint f(x)\,\id x$ is plotted  versus $\varepsilon$ in the inset.\\

\noindent {\bf Perturbative approach:}
We continue with the general set-up.
The equilibrium distribution $\rho_x^{\text{eq}}(\id \eta)$ satisfies the Gibbs formula
\bea
\nabla_x \rho_x^{\text{eq}}(\id \eta)
&=& - \rho_x^{\text{eq}}(\id \eta)\,[ \beta \nabla_x U(x,\eta)\, + \nabla_x \log Z_x ] \label{ere}
\eea
Multiplying the above relation  with $\rho_x(\id \eta)/\rho_x^{\text{eq}}(\id \eta),$ where  $\rho_x(\id \eta)$ is the stationary nonequilibrium density and integrating over $\eta$  we find the force $f(x) = \frac 1{\beta}\nabla_x \log Z_x + g(x)$ with nonequilibrium correction
\begin{equation}
\label{aha}
g(x) = -\frac 1{\beta}\,
\int \rho_x^{\text{eq}}(\id \eta)\,
\nabla_x  \frac{\rho_x(\id \eta)}{\rho_x^{\text{eq}}(\id \eta)}
\end{equation}
as an expectation with respect to the equilibrium reference.
The additional nonequilibrium force hence derives from the nonequilibrium correction $h_x$ in the stationary density,
$\rho_x(\id\eta) = \rho_x^\text{eq}(\id\eta)\,[1 + h_x(\eta)]$, as does the work (scalar product of force with displacement) performed on the probe,
\begin{equation}\label{pert}
g(x)\cdot \id x = -\frac{1}{\beta}\,\int
\rho_x^{\text{eq}}(\id \eta)\,h_{x+\id x}(\eta)
\end{equation}
No close-to-equilibrium assumptions have been made so far.  Yet, \eqref{aha}--\eqref{pert} invite building perturbation expansions $g(x) = \ve g_1(x) + \ve^2 g_2(x) + ... $  around the equilibrium case $\ve =0$. For such an expansion
we assume that the medium satisfies the local detailed balance condition~\cite{mn,tas}, which allows to quantify the amount of nonequilibrium driving in terms of a parameter $\varepsilon$ and through which the
(extra)
entropy flux to the thermal bath can physically be identified as for example the $s(\eta,\eta')$ in~\eqref{aa}.

The first-order nonequilibrium correction in the stationary distribution is given by the McLennan formula~\cite{mac,mcL},
which allows a thermodynamic characterization of $h_x(\eta)$.  More precisely, always in the linear regime, $h_x / \beta$ is given in terms of the excess work as has appeared in constructions of steady state thermodynamics \cite{sasatasaki,Oono}.
The notion of excess work arises from the fact that the nonequilibrium medium for fixed $x$ constantly dissipates work $W$ (as Joule heating) even at stationarity.  When now the probe position $x\rightarrow x +\id x$ changes, the medium starts to relax to a new stationary condition $\rho_x\rightarrow\rho_{x+\id x}$ with extra, indeed excess, work $\dbar W^\text{ex}(x)$  done by the  non-conservative forces; see Fig~\ref{fig:force}(b) for a representation. That is of course also dissipated as heat in the thermal environment. The McLennan scheme then applies to \eqref{pert} and readily implies that
the work done on the probe by the medium equals the excess work done on the medium by the nonequilibrium drive,
\begin{equation}\label{exc}
g_1(x) \cdot \id x = \dbar W^\text{ex}(x)
\end{equation}
That identity is the first result of the present Letter; it is not true in general beyond the linear regime.
The result can be extended in the same regime of weak nonequilibrium thermodynamics to include processes that are not isothermal using quasi-static energetics.\\

\noindent{\bf Kinetic aspects:}
As a new application of statistical forces we find that they also give information about the dynamical activity in the medium.  Just like excess work above, notions of dynamical activity while clearly relevant for nonequilibrium studies so far lack operational meaning.

\begin{figure}[th]
  \centering
  \includegraphics[width=8.8 cm]{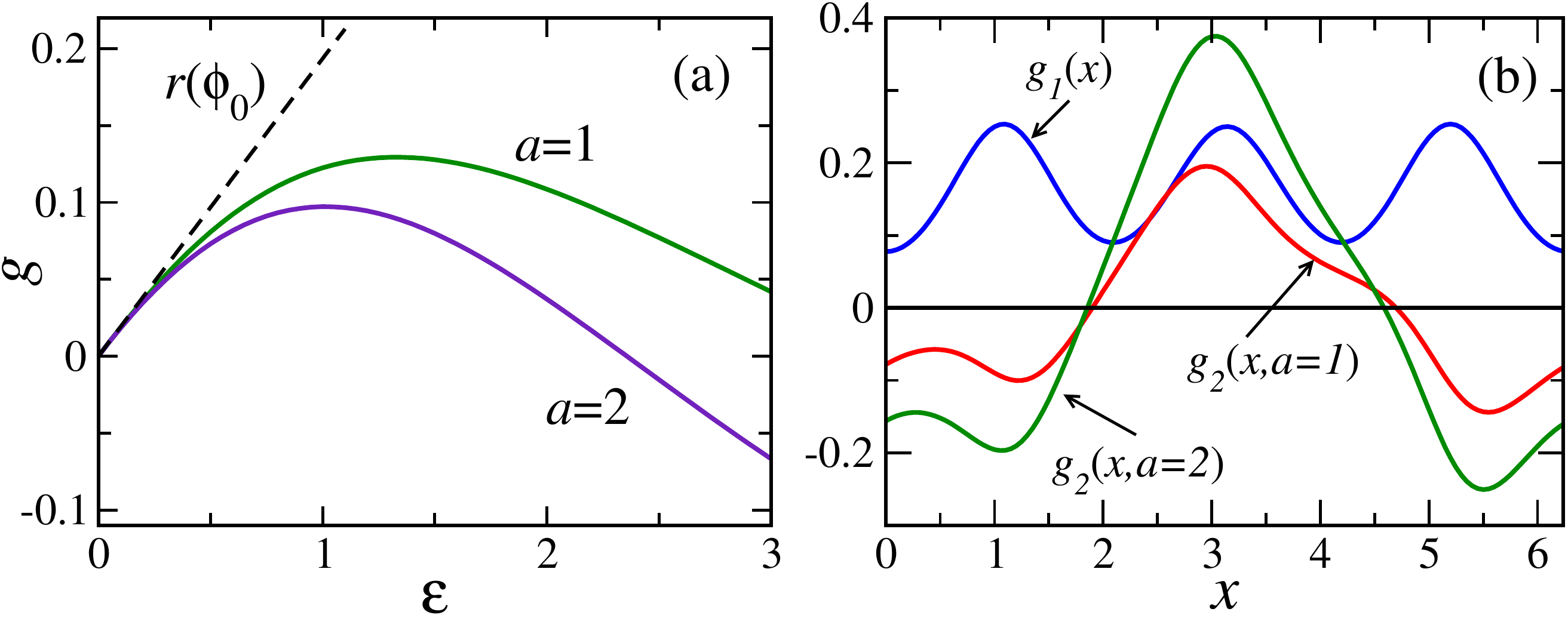}
  \caption{(Color online) (a)  The statistical force $g(x=\pi/2)$ as function of $\ve$.  The initial slope depends on the relative equilibrium reactivity $\phi_0$ (here $\phi_0=0.5$). For larger drive $\ve$ the dependence on the reactivity parameter $a$ becomes visible. (b) Linear $g_1(x)$ and second order correction $g_2(x)$ as a function of the probe position for $a=1,2$ and $\phi_0=1$. The linear order is independent of $a$ while the second order depends explicitly on it.} 
  \label{fig:fren}
\end{figure}

The first point is that we can infer from the linear order $g_1$
information on the equilibrium reactivities. We show it with the example in \eqref{aa}. The
relative
equilibrium frequency of undirected transitions between states is there governed via the $\phi_0$. While that $\phi_0$ is not influencing the equilibrium free energy, it does become visible to first order $g_1$ in the statistical force.  In other words, the relative dynamical activity in paths of equilibrium media can be found from applying a small force and from measuring the resulting mechanical effect on the probe. Fig.~\ref{fig:fren}(a) shows the dependence on $\phi_0$ in the slope  of the statistical force $g(x=\pi/2)$ at small $\ve$; $r(\phi_0)=g_1(\pi/2) $.  The more general dependence is
\[ 
g_1(x) = \frac{A_x + B_x \,\phi_0}{C_x + \phi_0}
\] 
where the $A,B,C$ are only depending on $x$ (and not on $a$ for example).

But there is more:  the second order is able to pick up the $\ve-$dependence (parameter $a$) in the  reactivities, invisible to linear order.
That is in line with the analysis of higher order effects in the response formalism in \cite{fren}.
The linear and second order corrections in the statistical force generated by~\eqref{aa} are plotted separately in Fig.~\ref{fig:fren}(b).
The statistical force starts to depend on the change  $a$ in reactivity  only from second order onwards around equilibrium. Alternatively, from the force on the walker we can detect information about the reactivity-change under the driving field.\\

\noindent{\bf Nonadditivity:}
The medium $\eta$ can consist of multiple reservoirs which are spatially separated or are only in weak contact.  We know that in the equilibrium case, because of the locality of the interaction or from the additivity of the free energy ${\cal F}(x),$  the statistical forces add --- the total force on the probe is the sum of the forces from the different reservoirs.    To be specific let us think of the situation where the probe is coupled to two different media both of which can be driven out of equilibrium in general. From equation \eqref{aha}, we see what is needed for additivity of statistical forces: the stationary $\eta-$distribution must approximately factorize over its constituents.  Or, the coupling with the second reservoir should not completely change the stationary distribution in the first reservoir. That is exactly what does not need to be the case for coupled nonequilibrium reservoirs, and thus can lead to nonadditivity of forces.

\begin{figure}[ht]
  \centering
  \includegraphics[width=3.75cm]{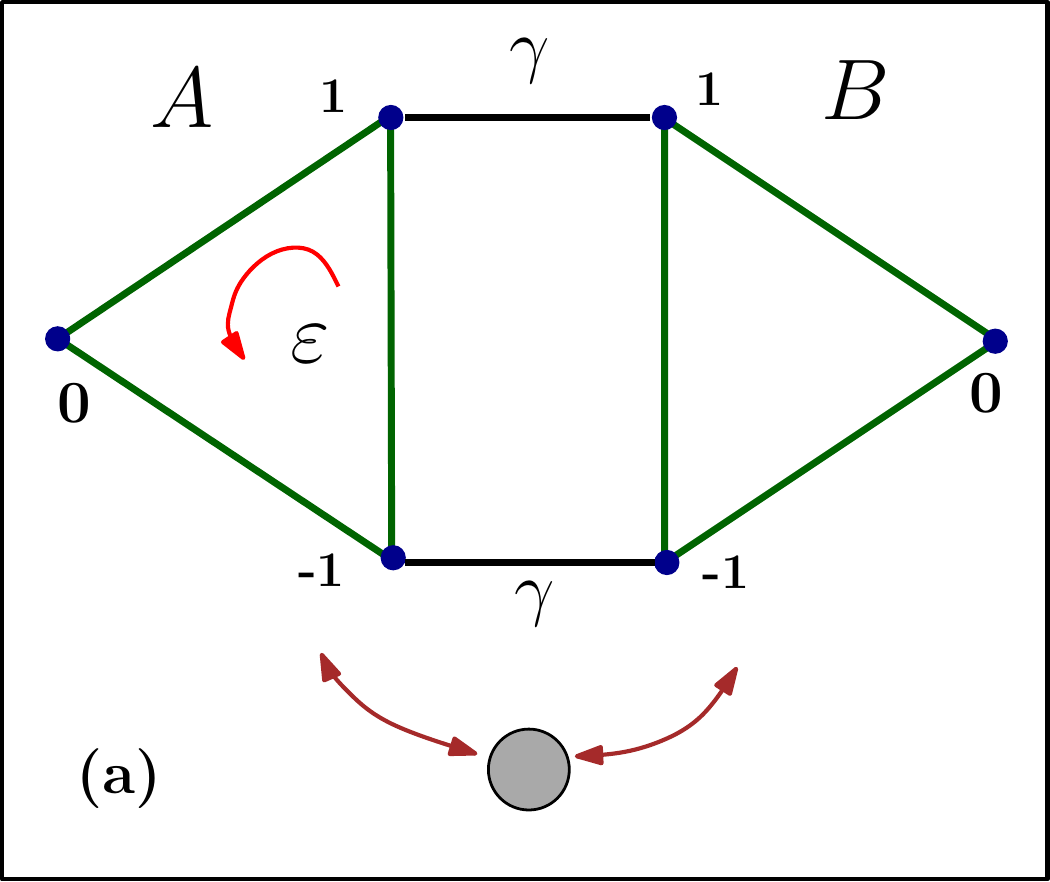}\hspace*{0.15cm}
  \includegraphics[width=4.15 cm]{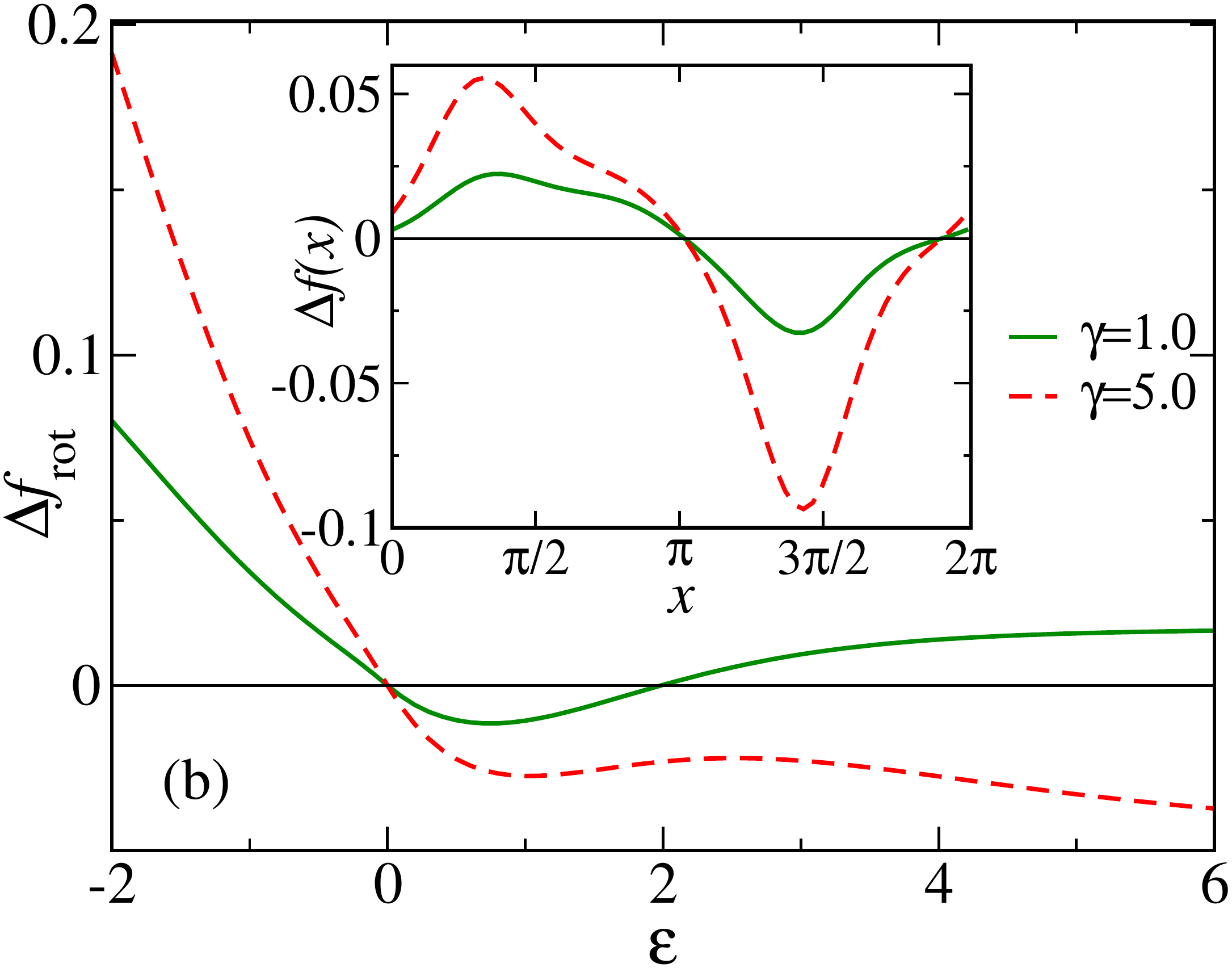}
  \caption{(Color online) (a) Coupling of two internal degrees of freedom to the probe.  The A-reservoir is driven and will create a current also in the B-reservoir, making the forces nonadditive. (b) Nonadditivity in the rotational component of the total force $\Delta f_\text{rot}$ {\it versus}  $\varepsilon$ for $\gamma=1.0,5.0$ and $a=2, \phi_0=1$. In the inset: the deviation from additivity $\Delta f(x)$ versus the probe position $x$ at $\ve=2.0.$}
 \label{fig:non_add_schm}
\end{figure}

In the simplest case we have the probe in contact with independent particles distributed over two parts, submedium A and submedium B. The total statistical force is then of the form
\begin{equation}\label{fpp}
f = \rho_A\, f_A + \rho_B\, f_B
\end{equation}
where $\rho_{A,B}$ are the concentrations and $f_{A,B}$ are the forces per particle.
There is a parameter $\gamma\geq 0$ that couples A and B, and there is a nonequilibrium driving $\ve$ that either drives the particles in A, or in B, or in both parts.  Particles can move between A and B as long as $\gamma >0$.  Both the concentrations $\rho^{\ve,\gamma}_{A,B}$ and the forces per particle $f^{\ve,\gamma}_{A,B}$ generally
depend on the driving $\ve$ as well as on the coupling $\gamma$. If the connection gets interrupted, setting suddenly $\gamma=0$, then we generally observe a shift in the statistical force, even if the concentrations remain unchanged. Indeed, by interrupting the flow between $A$ and $B$ the currents and the stationary distributions may change in both submedia, and the forces per particle take on new stationary values.
The observed change in the force or measure of nonadditivity is then defined as
\begin{equation}\label{nona}
\Delta f = \rho_A^{\ve,\gamma}\,[f^{\ve,\gamma}_{A} - f^{\ve,0}_{A}] + \rho_B^{\ve,\gamma}\,[f^{\ve,\gamma}_{B} - f^{\ve,0}_{B}]
\end{equation}

To illustrate the nonadditivity of nonequilibrium statistical forces we take again the example of a diffusive probe, now coupled to two reservoirs $\eta = \{\eta_A, \eta_B \}.$ For simplicity we again consider each of the reservoirs as a three state rotator with  $\eta=-1,0,1$ as before. The probe is coupled to the media through energy function  $U(x,\eta_\alpha) = \delta_{\alpha,A}[\eta \sin x+ 2\eta^2\cos x] + \delta_{\alpha,B}[\eta \cos x+ 2\eta^2\sin x]$ which is a sum of potentials similar to \eqref{eq:Ux}. Additionally the two reservoirs are also coupled to each other via  two distinct `bridges' at $1_A\leftrightarrow 1_B$ and $-1_A \leftrightarrow -1_B$ with jump rates
\bea
k^x(\pm 1_A, \pm 1_B) = \gamma\,
e^{- \frac \beta 2\,[ U(x,\pm 1_B) -U(x,\pm 1_A) ]}
\eea
and similarly for $B\rightarrow A$; see Fig.~\ref{fig:non_add_schm}(a) which represents a doubling of  internal degrees of freedom with respect to the situation depicted in Fig.~\ref{fig:setup}(b).
The whole system can be thought of as the probe being coupled to indistinguishable independent walkers hopping on the 6 states $\eta_\alpha$ with $\eta=-1,0,1$ and $\alpha=A,B$. Writing
$Z_{A,B} = \sum_{\eta \in A,B} e^{-\be U(x,\eta)}$  the equilibrium statistical force of the form \eqref{fpp} equals
\[ 
f_\text{eq}
= \frac{1}{\be}\nabla_x\log(Z_A + Z_B)
= \rho_A^0 f^0_A + \rho_B^0 f^0_B
\] 
where $f^0_{A,B} = \frac{1}{\be} \nabla_x \log Z_{A,B}$ is the mean force per particle from $A$ and $B$, respectively, and
$\rho^0_{A,B} =  Z_{A,B} / (Z_A + Z_B)$ is the probability of the particle to be in $A$ or $B.$
In equilibrium the statistical force is strictly additive because under detailed balance the prefactor $\gamma$ does not change the Boltzmann occupation statistics; the $f^0_{A,B}$ do not depend on the coupling $\gamma$ so that by cutting the connection between $A$ and $B$ nothing is changed. However, when either $A$ or $B$ (or both) are driven out of equilibrium the bridges and the transition strength $\gamma$ become important. For example, a drive in medium $A$ typically induces a current in $B$ and changes the stationary distribution, thus giving rise to a different statistical force from $B$.  The nonadditivity \eqref{nona} is shown for that example in Fig.~\ref{fig:non_add_schm}(b). The rotational component of the difference $\Delta f$ is plotted there for different $\gamma$.  The difference $\Delta f(x)$  for a fixed driving field $\varepsilon$  in A as in \eqref{aa} is shown in the inset. As expected  $| \Delta f|$ increases with the strength of the coupling $\gamma.$

The fact that nonadditivity already occurs for noninteracting particles is perhaps surprising. It is however directly related to the phenomenon that
local changes
in systems with conserved quantities
can have global effects on the  nonequilibrium stationary distribution, \cite{bid}.

For interacting particles in the media with generic long range correlations, that effect can only be enhanced and other issues become also important.  Indeed, as a related effect for spatially extended systems, the difference between bulk and boundary contributions to statistical forces is  fading \cite{Kafri15}.  All that is relevant for the status and possibility of steady state thermodynamics, the macroscopic theory of nonequilibrium systems.  In particular, the difference between bulk versus surface effects is important in the usual thermodynamics, and the equations of state typically connect pressure with bulk properties.  The above  simple arguments and example already
indicate possible limitations for macroscopic nonequilibria.\\

\noindent {\bf Conclusion:}
Statistical forces on quasi-static probes in nonequilibrium media have a number of new interesting features with respect to the equilibrium case.  In linear order around equilibrium they can still be obtained from irreversible thermodynamics, where the force measures the excess work
of driving forces on the medium.  Note that already there the forces can be rotational and nonadditive as a result of long range effects.
For non-thermodynamic information these forces can be used to measure relative and excess dynamical activities in the medium, a concept of growing importance for which little operational tools are available so far.    Nonequilibrium conditions 
might affect reactivities and time-symmetric aspects of the medium, and they are present in the statistical force on the probe from second order onwards. Such higher order effects are due to the coupling of the entropy flux with excess dynamical activity, \cite{fren}.
That also contributes to the violation of the second fluctuation--dissipation relation in generalized Langevin equations for a probe coupled to nonequilibrium media,  \cite{stef}:  there again the friction picks up more kinetic information from the
medium.  Statistical forces thus provide an important tool to probe the nature and properties of
nonequilibrium systems.


\end{document}